\documentclass[prd, aps, superscriptaddress, preprintnumbers, twocolumn, floatfix, nofootinbib]{revtex4-2}
\pdfoutput=1

\usepackage{amsfonts}
\usepackage{amsmath}
\usepackage{amssymb}
\usepackage{bm}
\usepackage{dcolumn}
\usepackage{graphicx}
\usepackage[latin1]{inputenc}
\usepackage{latexsym}
\usepackage{rotating}
\usepackage{hyperref}
\usepackage{graphicx}
\usepackage{xcolor}

\newcommand\be{\begin{equation}}
\newcommand\bea{\begin{eqnarray}}
\newcommand\ee{\end{equation}}
\newcommand\eea{\end{eqnarray}}

\renewcommand{\d}{{\mathrm{d}}}
\renewcommand{\[}{\left[}
\renewcommand{\]}{\right]}
\renewcommand{\(}{\left(}
\renewcommand{\)}{\right)}

\newcommand{\Mpl}{M_{\textrm{Pl}}}

\def\doi{http://doi.org}

\def\d{\mathrm{d}}

\begin{document}

\title{BFSS Matrix Model Cosmology: Progress and Challenges}

\author{Suddhasattwa Brahma}
\email{suddhasattwa.brahma@gmail.com}
\affiliation{Higgs Centre for Theoretical Physics, School of Physics \& Astronomy,
University of Edinburgh, Edinburgh EH9 3FD, UK.}

\author{Robert Brandenberger}
\email{rhb@physics.mcgill.ca}
\affiliation{Department of Physics, McGill University, Montr\'{e}al, QC, H3A 2T8, Canada}

\author{Samuel Laliberte}
\email{samuel.laliberte@mail.mcgill.ca}
\affiliation{Department of Physics, McGill University, Montr\'{e}al, QC, H3A 2T8, Canada}

\date{\today}

\begin{abstract} 

We review a proposal to obtain an emergent metric space-time and an emergent early universe cosmology from the BFSS matrix model. Some challenges and directions for future research are outlined.

\end{abstract}

\maketitle

\section{Introduction}

Current models of the very early universe are generally based on effective field theory (EFT), i.e.  on considering matter fields minimally coupled to Einstein gravity.  This is true both for the inflationary scenario \cite{Guth} and for most alternatives to it (see e.g. \cite{RHBaltrev} for a comparative review of alternatives to inflation). However, there are conceptual challenges which effective field theory descriptions of cosmology face. For example,  in the case of scalar field matter - ubiquitously used in early universe models - the range of such fields \cite{distance} and and the shape of \cite{dS} of their potentials are constrained if we demand that the effective fields descends from some fundamental physical theory (the best candidate of which is superstring theory).  Independent of superstring theory, the {\it Trans-Planckian Censorship Conjecture} (TCC) \cite{TCC1} severely constrains inflationary models which are based  on an EFT analysss \cite{TCC2} \footnote{See e.g. \cite{TCC3} for discussions of some arguments in support of the TCC. Note that scenarios of inflation which are not based on EFT analyses (see \cite{Dvali, Keshav} for examples) are not necessarily subject to these constraints.}. 

An EFT analysis of matter also leads to the famous cosmological constant problem: considering matter fields coupled to Einstein gravity, and canonically quantizing the fields by expanding them in plane wave modes and quantizing each mode like a harmonic oscillator, then the vacuum energy of all of these modes yields a contribution to the cosmological constant which is many orders of magnitude (120 orders in a non-supersymmetric model) larger than what is compatible with current observations \footnote{To avoid the Planck catastrophe, an ultraviolet (UV) cutoff scale must be imposed, the value of which determines the predicted cosmological constant.}.

The abovementioned problems indicate that one needs to go beyond point particle-based EFT to obtain a consistent picture of cosmology.  Here, we will review an approach \cite{us1, us2} to obtain a model of the very early universe which is not based on EFT. Our starting point is the BFSS matrix model \cite{BFSS}, a proposed non-perturbative definition of superstring theory. (see also \cite{Hoppe} for early work on the quantum theory of membranes, and \cite{different} for other matrix models related to superstring theory) We outline an avenue of how to obtain an emergent space-time, an emergent metric, and an emergent early universe cosmology.  We consider the matrix model to be in a thermal state. This leads to density fluctuations and gravitational waves on the emergent space-time, and we indicate how the resulting power spectra of these fluctuations are scale-invariant,  and consistent with current observations.

In this context, we remind the reader that the inflationary scenario \cite{Guth} is not the only early universe scenario which is consistent with current cosmological observations (see e.g. \cite{RHBaltrev} for comparative reviews of inflation and alternatives). An alternate scenario is the {\it emergent scenario} which is based on the assumption that there is a primordial phase which can be modelled as quasi-static (from the point of view of the Einstein frame), and which then undergoes a phase transition to the Standard Big Bang phase of expansion.  

A model for an emergent scenario is {\it String Gas Cosmology} (SGC) \cite{BV} (see also \cite{Perlt} for earlier work and \cite{SGCrev} for a review) where it is assumed that the universe begins in a quasi-static {\it Hagedorn} phase, a phase in which a gas of closed superstrings is in thermal equilibrium at a high temperature close to the limiting Hagedorn value \cite{Hagedorn}.  The initial size of all of the spatial dimensions is taken to be the same.  However, a dimension of space can only expand if the winding modes in that direction disappear,  and, since strings are two-dimensional world sheets, the intersection of winding strings (which is necessary for winding modes to disappear) is not possible in more than four large space-time dimensions \footnote{In this argument it is assumed that there are no long range attractive forces between winding strings. However, since the local curvature perpendicular to gauge strings vanishes \cite{Vilenkin}, it is reasonable to make this assumption.}.  This gives a nice way to understand why there are only three of the nine spatial dimensions that superstring theory predicts can become large.  It can then be shown \cite{SGCperts} that thermal fluctuations of this string gas lead to scale-invariant spectra of cosmological perturbations and gravitational waves, with a slight red tilt for the curvature fluctuations and a slight blue tilt for the gravity wave spectrum.

What was missing in SGC is a consistent dynamical framework for analyzing the cosmological evolution. To obtain such a framework, we need to go beyond an EFT analysis, and this is where our work comes in \footnote{See also \cite{Vafa} for a different approach to obtaining an emergent cosmology from superstring theory which matches well with the SGC model.}.  

Note that, as we will review below, the computation of cosmological perturbations and gravitational waves in our BFSS model is based on the same formalism used in SGC, with very similar results.

In the following section we will present a brief review of the BFSS matrix model. In Section (\ref{secthree}) we then present our proposal for emergent time, space and metric from the matrix model. On this basis,  in Section (\ref{secfour}) we develop a new picture of an emergent universe and compute the curvature fluctuations and gravitational waves which are predicted if one starts with a high temperature state of the BFSS matrix model. We use natural units in which the speed of light and Planck's constant are set to one.

\section{BFSS Matrix Model} \label{sectwo}

Starting point is the BFSS matrix model \cite{BFSS}, a quantum theory involving ten bosonic time-dependent $N \times N$ Hermitean matrices $A_0(\tau)$ and $X_i,(\tau), i = 1 \cdots 9$ and their sixteen Fermionic superpartners $\theta_{a, b}(\tau)$ (also $N \times N$ matrices) which transform as spinors under the $SO(9)$.  The Lagrangian is 
\bea \label{BFSSaction}
L \, &=& \, \frac{1}{2 g^2} \bigl[ {\rm Tr} \left \{\frac{1}{2} \(D_{\tau} X_i\)^2 - \frac{1}{4} \[X_i, X_j\]^2\right\} \nonumber \\
&& - \theta^T D_{\tau} \theta - \theta^T \gamma_i \[ \theta, X^i \] \bigr] \, ,
\eea
where the covariant derivate involves the temporal bosonic matrix $A_0$
\be
D_{\tau}:=\partial_{\tau} - i\[A_0(\tau), \cdot\] \, .
\ee
 In the above, $\tau$ is the BFSS time, and $\gamma_i$ are Clifford algebra matrices (see e.g. \cite{Ydri, Taylor} for reviews of the BFSS matrix model).
 
This Lagrangian has both a $U(N)$ symmetry and a $SO(9)$ symmetry under rotations of the spatial matrices (and the corresponding transformation of the spinors).  The partition function of the theory is given by the functional integral
\be
Z \, = \, \int dA d\theta e^{iS_{BFSS}} \, ,
\ee
where $S_{BFSS}$ is the action obtained by integrating the Lagrangian over time, and the integration measures are the Haar measure for the bosonic variables and the spinorial counterpart.

This matrix model was introduced \cite{BFSS} as a candidate for a non-perturbative definition of superstring theory (more precisely of M-theory).  String theory emerges in the limit $N \rightarrow \infty$ with $\lambda \equiv g^2 N$ held fixed.

We will consider this matrix model in a high temperture state, the temperature being denoted by $T$. By this assumption, there is no evolution of the system in BFSS time $\tau$. However,  below we will extract an emergent time $t$ (which is different from the time $\tau$) in terms of which there is non-trivial dynamics. 

In our high temperature state,  the spatial matrices $X_i$ are periodic in Euclidean BFSS time $\tau_E$ and we can expand them in Matsubara modes
\begin{equation}
X_i(\tau_E) \, = \, \sum_n X_i^n e^{2 \pi n T \tau_E} \, ,
\end{equation}
where $n$ runs over the integers \footnote{Note that this integer $n$ is unrelated to the integers $n_i$ used in later sections to introduce comoving coordinates.}.  At high temperatures, the BFSS action is dominated by the $n=0$ modes
\be
S _{BFSS} \, = \, S_{IKKT} + {\cal{O}}(1/T) \, ,
\ee
where $S_{IKKT}$ is the bosonic part of the action of the IKKT matrix model \cite{IKKT}, which, in terms of rescaled matrices $A_i$
\begin{equation}
A_i \, \equiv \, T^{-1/4} X_i^0  
\end{equation}
is given by
\be \label{IKKTaction}
	S \, = \,  -\frac{1}{g^2} \text{Tr}\(\frac{1}{4} \[A^a, A^b\]\[A_a,A_b\] \) \, .
\ee

In the following we will use the matrices $A_0$ and $A_i$ (i.e. the zero modes of the BFSS matrices $A_0$ and $X_i$) to extract an emergent space-time. The $n \neq 0$ modes of the matrices will play an important role when obtaining cosmological fluctuations and gravitational waves.

\section{Emergent Metric Space-Time} \label{secthree}

Without loss of generality we can work in a basis in which the temporal matrix $A_0$ is diagonal. The expection values of the diagonal elements are identified with the {\it emergent time}.  We will order the diagonal elements in increasing value
\be
A_0 \, = \, {\rm{diag}} (t_1, ... , t_N) \, ,
\ee
with $t_i \leq t_j$ for $i < j$.

Numerical simulations of the IKKT model \cite{IKKT-time} indicate that
\be
\frac{1}{N} \bigl< {\rm Tr} A_0^2 \bigr> \sim \kappa N \, ,
 \ee
 where $\kappa$ is a (small) constant. Assuming that the eigenvalues are separated equidistantly, it then follows that the maximal eigenvalue (the maximal value of the emergent time) scales as
\be
t_{{\rm{max}}} \,  \sim \, \sqrt{N} \, .
\ee
and that hence the difference between the discrete values of time scales as
 \be \label{Deltat}
 \Delta t \, \sim \frac{1}{\sqrt{N}} \, .
\ee
Thus, in the $N \rightarrow \infty$ limit we obtain an emergent continuous and infinite time. Since the eigenvalues are symmetric about $t = 0$, the time that emerges is infinite both in the past and in the future. Hence, there is neither a Big Bang nor a Big Crunch singularity.

\begin{figure}
    \includegraphics[scale = 0.35]{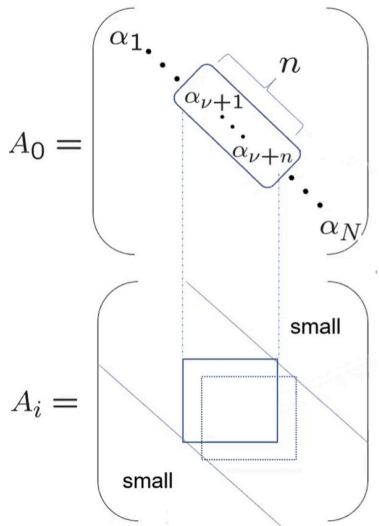}
    \caption{In the basis in which the $A_0$ matrix is diagonal with the diagonal elements defining emergent time,  we can use spatial $n_i \times n_i$ submatrices ${\bar{A_i}}(t)$ located a distance $t$ down the diagonal to define the i'th spatial direction at time $t$.  This figure is taken from \cite{Ydri} with permission.} 
    \label{matrix}
\end{figure}

Turning to the spatial matrices, we follow the proposal of \cite{IKKT-space} and consider, for each integer $m$,  $n_i \times n_i$ (where $n_i$ is an integer sufficiently smaller than $N$) spatial submatrices ${\bar{A_i}(t(m))}$ centered a distance $m$ down the diagonal (see Figure 1)
\be
({\bar{A_i}})_{I, J}(t(m)) \, \equiv \, (A_i)_{m + I, m + J} \, .
\ee
where $t(m)$ is the m'th temporal eigenvalue.  Following \cite{IKKT-space} we can define the {\it extent of space} parameters $x_i(t)$ in the i'th spatial direction at the time $t$ via
\be
x_i(t)^2 \, \equiv \, \left\langle \frac{1}{n} \text{Tr} ({\bar{A_i}})(t))^2 \right\rangle \, .
\ee
With this definition of {\it emergent space}, there is non-trivial time evolution in terms of the emergent time.

In the IKKT model, it has been observed via numerical simulations that, in the presence of the fermionic terms in the action, the state that minimizes the free energy breaks the $SO(9)$ symmetry into $SO(6) \times SO(3)$. Of the nine extent of space parameters, six remaiin small and three increase \cite{IKKT-PT}. This result can also be confirmed by means of Gaussian expansion calculations \cite{Lowe}.  The emergence of three macroscopic spatial dimensions with the other six remaining microscopic reminds one of the same phenomenon which was discovered in the context of String Gas Cosmology \cite{BV}, where string winding modes prevent spatial dimensions from expanding.  In order to obtain large dimensions, string winding modes about these dimensions must annihilate, and this is not possible in more than three large spatial dimensions since string winding modes will have vanishing intersection probability (see \cite{Subodh} for a more detailed analysis).

Very recently, we have been able to show \cite{us3} that in the presence of the fermionic terms, the $SO(9)$ symmetry will also be spontaneously broken in the BFSS matrix model.  This result was derived using a Gaussian expansion calculation. With this tool, it is not yet possible to determine the symmetry of the state which minimizes the free energy, but the intuition gained from String Gas Cosmology leads us to expect that the symmetry breaking pattern will be the same as what was observed in the IKKT model.

Making use of the Riemann-Lebesgue Lemma,  it can be shown from the bosonic IKKT action that the off-diagonal elements of the spatial matrices decay once one departs beyond a critical distance $n_c$ from the diagonal:
\be
\sum_i \bigl< |A_i|^2_{ab} \bigr> \, \rightarrow \, 0 \,\, {\rm{for}} \,\, |a - b| > n_c
\ee
where
\be
n_c \, \sim \, \sqrt{N}
\ee
This result can be confirmed by numerical simulations (see \cite{IKKT-recent} for recent work on the IKKT matrix model) which also indicate that
\be \label{numresult}
\sum_i \bigl< |A_i|^2_{ab} \bigr> \, \sim \ {\rm{constant}} \,\, {\rm{when}} \,\, |a - b| < n_c \, ,
\ee
a result for which we at present do not have any analytical evidence.

Our proposal \cite{us2} is to view $n_i$ as the i'th comoving spatial coordinate,  and
\be
l_{phys, i}^2(n_i,  t) \, \equiv \, \left\langle \text{Tr} ({\bar{A_i}})(t))^2 \right\rangle \, ,
\ee
as the physical length square of the curve from $n _i = 0$ to $n_i$  along the i'th coordinate direction.  It then follows from (\ref{numresult}) that
\be
l_{phys, i}(n_i) \, \sim \, n_i  \,\, {\rm{for}} \,\, n_i < n_c \, .
\ee
Given the definition of length and comoving coordinates we can extract the spatial metric component $g_{ii}$ in the usual way:
\be
g_{ii}(n_i)^{1/2} \, = \, \frac{d}{dn_i} l_{phys, i}(n_i) 
\ee
and we obtain
\be
g_{ii}(n_i, t) \, = \, {\cal{A}}(t) \delta_{ii} \,\,\,   i = 1, 2, 3  \, ,
\ee
where ${\cal{A}}(t)$ is an increasing function of $t$ at late times. Making use of the remnant $SO(3)$ symmetry we obtain
\be \label{spatial}
g_{ij}({\bf{n}}, t) \, = \, {\cal{A}}(t) \delta_{ij} \,\,\,   i, j = 1, 2, 3  \, ,
\ee
where ${\bf{n}}$ stands for the comoving 3-vector corresponding to the three large dimensions.

We have thus indicated how one may obtain emergent time, space and spatial metric from the zero modes of the BFSS matrix model. The emergent space-time is infinite both in temporal and spatial directions, and from (\ref{spatial}) it follows that the metric which emerges on the emergent large three-dimensional space is spatially flat.  Hence, in the scenario which we are proposing the famous horizon and flatness problems of Standard Big Bang cosmology are automatically solved, without the need for any phase of inflation \cite{Guth} or superslow contraction \cite{Ekp}. In the following section we will show that the thermal fluctuations in the BFSS state automatically lead to approximately scale-invariant spectra for cosmological perturbations and gravitational waves.

\section{Emergent Early Universe Cosmology} \label{secfour}

The cosmology which emerges from the BFSS matrix model is an {\it emergent cosmology}. The classical notion of space-time only makes sense at late emergent time . Solving the classical bosonic matrix equations for the three large spatial dimensions at late times yields \cite{IKKT-late}
\be
{\cal{A}}(t) \, \sim \, t^{1/2} \, , 
\ee
i.e. expansion like in a radiation-dominated universe.  We expect that the inclusion of the fermionic sector will change this scaling and produce a matter component which scales as pressureless dust.  Thus, in our cosmology there is a direct transition from the emergent phase to that of Standard Big Bang cosmology, like in String Gas Cosmology. Note that there is no sign of a cosmological constant.  Thus, it appears that the cosmology which emerges from the BFSS matrix model will be free of the infamous cosmological constant problem. On the other hand, it is at the present time unclear how Dark Energy will emerge.  Based on the Swampland Criteria \cite{distance, dS} and the ``Trans-Planckian Censorship Conjecture'' \cite{TCC1} it follows that Dark Energy cannot be a cosmological constant. It may be a Quintessence field \cite{Quin} which emerges from the matter sector (see e.g. \cite{BBF} for an attempt to obtain a unified dark sector (Dark Matter and Dark Energy) from string theory).

Since we are considering the BFSS matrix model in a high temperature state, there will be thermal fluctuations, and these will lead to curvature fluctuations and gravitational waves like in String Gas Cosmology (SGC) (see \cite{SGCperts} for analyses of the generation of cosmological perturbations and gravitational waves in SGC).

At late times, the classical metric of the $3+1$ dimensional space-time can be written at the level of linearized fluctuations as (see e,g, \cite{MFB, RHBfluctsrev} for reviews of the theory of cosmological perturbations)
\begin{equation}
ds^2 \, = \, a^2(\eta) \bigl( (1 + 2 \Phi)d\eta^2 - [(1 - 
2 \Phi)\delta_{ij} + h_{ij}]d x^i d x^j\bigr) \,. 
\end{equation}
where $\Phi$ (which depends on space and time) is the represents the curvature fluctuations and the transverse tracelss tensor $h_{ij}$ stands for the gravitational waves \footnote{In writing the perturbed metric in this form we have chosen to work in longitudinal gauge.}.

According to the linearized Einstein equations, the curvature fluctuations of the geometry are determined in terms of the energy density perturbations via
\begin{equation}  
k^{-4} \langle\delta T^0{}_0(k) \delta T^0{}_0(k)\rangle \, , 
\end{equation}
and the gravitational waves are given by the off-diagonal pressure fluctuations
\begin{equation} 
\label{tensorexp} \langle|h(k)|^2\rangle \, = \, 16 \pi^2 G^2 
k^{-4} \langle\delta T^i{}_j(k) \delta T^i{}_j(k)\rangle \,\,\, (i \neq j) \, .
\end{equation}
Since the energy density perturbations in a region of radius $R$ are determined by the specific heat capacity $C_V(R)$ on that scale, the power spectrum of curvature fluctuations is given by
\begin{equation}
P(k) = k^3 (\delta \Phi(k))^2 = 16 \pi^2 G^2 k^{2}  T^2 C_V(R) \, ,
\end{equation}
where $k$ is the momentum scale associated with $R$.

For thermal fluctuations,  the matter correlation functions are given by taking partial derivatives of the finite temperature partition function. Since the partition function of the BFSS matrix model is our starting point, we are able to compute the observables. Specifically, the specific heat capacity is related to the internal energy $E(R)$ via
\begin{equation}
C_V(R) \, = \, \frac{\partial}{\partial T} E(R) \, ,
\end{equation}   
and the internal energy is given by the derivative of the partition function with respect to the inverse temperature $\beta$:
\begin{equation}
E \, = \, - \frac{\partial}{\partial \beta} {\rm{ln}} Z(\beta) \, .
\end{equation}
The computation of the internal energy $E(R)$ were carried out in \cite{us1}, based on work done in \cite{IKKT-TD} in the context of the IKKT matrix model.  We find
\be
E^2 \, = \, N^2 < {\cal{E}}>_{BFSS} 
\ee
with
\be
{\cal{E}} \, =  \,  - \frac{3}{4 N \beta} \int_0^{\beta} dt {\rm{Tr}}([X_i ,  X_j]^2) \, .
\ee
Evaluating ${\cal{E}}$ we obtain
\begin{equation} \label{result}
 E^2 \, = \, \frac{3}{4} N^2 \chi_2 T - \frac{3}{4} N^4 \alpha \chi_1 T^{-1/2} \, ,
 \end{equation}
 where $\chi_2$ is a constant while $\chi_1$ depends on $R$ via
 \begin{equation}
 \chi_1 \, = \, <R^2>_{BFSS} T^{-1/2} \, .
 \end{equation}
 The first term in (\ref{result}) yields a Poisson spectrum which dominates on ultraviolet scales, while the second term yields a scale-invariant contribution to the curvature power spectrum which dominates on the infrared scales relevant to current cosmological observations.
 
\section{Challenges and Future Directions}

 To summarize, we have presented a proposal to obtain time, space and a metric as emergent phenomena from the BFSS matrix model, taken in a high temperature state.  In the $N \rightarrow \infty$ limit, the time which emerges is infinite in both directions, and continuous.  Similarly,  the emergent space as infinite extent.  Given our proposal, the background metric which emerges is spatially flat and isotropic. Since we are considering a thermal state of the BFSS model, there are thermal fluctuations, and we have reviewed here how these perturbations yield scale-invariant spectra of curvature fluctuations and gravitational waves. Hence, our cosmological model solves the famous horizon and flatness problems of Standard Cosmology, and provides a causal mechanism for the origin of scale-invariant spectra of curvature fluctuations and gravitational waves,  all without the need to postulate a phase of cosmological inflation. The cosmology is non-singular, and the cosmological constant problem is absent.
 
 The $SO(9)$ symmetry of the BFSS action is spontaneously broken. In the case of the IKKT model, the state which minimizes the energy has $SO(6) \times SO(3)$ symmetry, with the extent of space increasing only in three spatial dimensions.  Based on solving the classical matrix equations, it appears that the cosmological expansion of the three large spatial dimensions is as in a radiation-dominated universe.  The presence of fermions plays a key role in obtaining this symmetry breaking. On the other hand, the fermionic sector has not been included in studying the late time dynamics of the three large spatial dimensions. An important open problem is to include the fermionic sector. A careful study of the full BFSS matrix model may also shed some light on the nature of Dark Energy.
 
An important open problem is to study the dynamics of this phase transition and to determine the key physical reason for the distinguished symmetry breaking pattern. We speculate that the reason may be related to the reason why three dimensions become large in SGC.  Once the dynamics of the phase transition is understood, we may be able to make predictions for the tilts $n_s$ and $n_t$ of the spectra of cosmological perturbations and gravitational waves, or at least determine consistency relations between the tilts.  We may be able to recover the same consistency relation $n_t = 1 - n_s$ as in SGC (and as in the recently proposed Ekpyrotic scenario mediated by an S-brane \cite{Ziwei}).

While our scenario yields emergent space and time, and an emergent metric,  it is important to investigate what effective action describes the evolution of localized matrix excitations. At low energies, we need to be able to recover Einstein gravity coupled to regular matter and radiation.

In conclusion, although we have indicated a promising approach to the emergence of space and time, and of an early universe cosmology, a lot of work needs to be done before being able to make successful contact with testable low energy physics.

\section*{Acknowledgments}

SB is supported in part by the Higgs Fellowship. SL is supported in part by FRQNT. The research at McGill is supported in part by funds from NSERC and from the Canada Research Chair program.  We thank B. Ydri for permission to use the figure (\ref{matrix}) which is taken from \cite{Ydri}.

\end{document}